\documentclass[journal]{IEEEtran}

\usepackage{amsmath,amsfonts,graphicx}
\usepackage{cite}
\usepackage{hyperref}
\hypersetup{colorlinks=true,linkcolor=blue,citecolor=blue,urlcolor=blue}

\usepackage{booktabs}

\usepackage{comment}

\usepackage{algorithm}
\usepackage{algpseudocode}
\usepackage{amsmath}

\title{Can We Reliably Predict the Fed’s Next Move? A Multi-Modal Approach to U.S. Monetary Policy Forecasting}

\author{Fiona Xiao Jingyi and Lili Liu*, National University of Singapore}

\begin{document}
	
	\maketitle
	
	\begin{abstract}
		Forecasting central bank policy decisions remains a critical challenge for financial institutions, investors, and policymakers, given the profound influence of monetary actions on market dynamics and macroeconomic conditions. In particular, anticipating changes in the U.S. federal funds rate is essential for effective risk management and the formulation of informed trading strategies. Yet, traditional approaches relying solely on structured economic indicators may fall short in capturing the forward-looking nuances conveyed through central bank communications.
		
		This study examines whether predictive performance can be improved by integrating structured macroeconomic data with unstructured textual signals from Federal Reserve communications. We adopt a multi-modal modeling framework to compare the effectiveness of traditional machine learning classifiers, transformer-based language models, and deep learning architectures across both unimodal and hybrid configurations.
		
		Empirical results demonstrate that hybrid models consistently outperform unimodal baselines. The strongest performance is achieved by integrating TF-IDF representations of FOMC texts with structured economic features in a gradient boosting (XGBoost) classifier, achieving a test AUC of 0.83. In contrast, deep learning approaches using FinBERT-derived sentiment probabilities provide marginal improvements in ranking but underperform in classification accuracy, particularly in the presence of class imbalance. SHAP analysis further reveals that interpretable, sparse features better reflect policy-relevant signals in formal monetary texts.
		
		These findings suggest that, in the context of financial policy prediction, simplicity and interpretability can be powerful assets. Our results offer practical insights for scholars, policymakers, and practitioners alike, highlighting the value of hybrid, transparent models for navigating the evolving landscape of central bank decision-making.
	\end{abstract}
	
	
	\section{Introduction}
	
	Forecasting interest rate decisions by the U.S. Federal Reserve is a core challenge in financial and macroeconomic analysis. These decisions influence asset prices, guide investor expectations, and underpin overall economic stability. Historically, such forecasts have relied heavily on structured macroeconomic indicators—such as inflation rates, employment levels, and GDP growth—closely aligned with the Fed’s dual mandate of price stability and full employment.
	
	In the aftermath of the Global Financial Crisis (GFC), the Federal Reserve has placed growing emphasis on forward guidance as a key component of monetary policy communication~\cite{cecchetti2020monetary, blinder2008central}. Formal channels—such as statements, speeches, meeting minutes, and press conferences—now serve not merely as reflections of policy decisions but as instruments to shape expectations and convey strategic intent~\cite{fortes2020tracking}. This development reflects a broader methodological shift: from purely rule-based, data-driven forecasting toward frameworks that integrate both structured economic indicators and unstructured textual signals.

	This development raises an important research question:
	
	\begin{quote}
		\textit{Can combining macroeconomic indicators with unstructured Fed communication improve the predictive accuracy and interpretability of interest rate forecasts?}
	\end{quote}
	
\begin{figure}[t]
		\centering
		\includegraphics[width=0.45\textwidth]{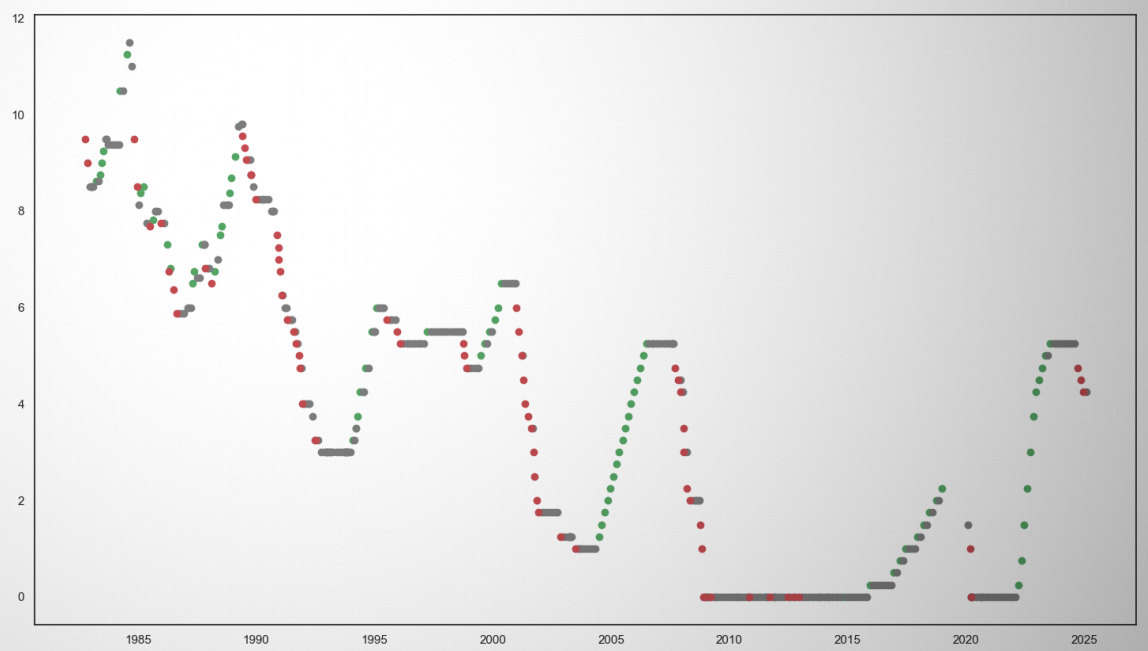}
		\caption{Federal Funds Rate trajectory (Jan 1980 – Jan 2025).}
		\label{fig:fed_rate_trend}
	\end{figure}

	As shown in Figure~\ref{fig:fed_rate_trend}, U.S. interest rates have exhibited marked volatility over the past decades, driven by complex macroeconomic and geopolitical dynamics. Anticipating these movements remains a high-stakes task for market participants and policymakers alike.
	
	To address this challenge, we introduce a multi-modal forecasting framework that integrates structured macroeconomic indicators with sentiment signals derived from Federal Reserve communications. Our main contributions are as follows:
	
	\begin{itemize}
		\item We develop a hybrid model that synthesizes sentiment probabilities with curated economic features to capture both quantitative fundamentals and qualitative policy narratives.
		\item We perform a rigorous comparative evaluation across multiple model architectures, quantifying the individual and combined predictive strengths of each data modality.
		\item We apply SHAP-based interpretability to clarify the role of textual and numerical features, enhancing transparency and supporting meaningful economic insight.
	\end{itemize}
	
	By jointly modeling structured and unstructured inputs, our approach advances monetary policy forecasting in both accuracy and interpretability. The proposed system achieves a test AUC of 0.83, while offering decision-relevant explanations aligned with the practical needs of economists and central bank observers.

	\section{Literature Review}
	
	The majority of existing research has focused on one data modality in isolation—either structured or unstructured. This has limited the potential to uncover richer insights from the interplay of data types. In what follows, we provide a structured review across both dimensions and highlight recent efforts toward multi-modal approaches.
	
	\subsection{Structured Economic Indicators in Policy Forecasting}
	Rule-based models, such as the Taylor Rule~\cite{taylor1993discretion}, represent early structured approaches to policy forecasting. These models relate policy rates to deviations in inflation and output gaps. Building on these foundations, econometric frameworks like vector autoregressions (VARs)\cite{apel2012monetary} have been widely used to capture dynamic interdependencies among macroeconomic variables. Furthermore, GARCH-type models\cite{bollerslev1986garch} have offered valuable tools for estimating and forecasting volatility in financial time series. Despite their rigor, these methods often lack the ability to incorporate qualitative or forward-looking signals embedded in central bank communication.
	
	\subsection{Central Bank Communications and Textual Analysis}
	In parallel, a growing body of work investigates the predictive utility of central bank discourse. Early efforts used basic textual metrics or manual annotations to assess tone and emphasis. The use of natural language processing (NLP) has since expanded, allowing for more systematic analysis. Texts such as meeting minutes, statements, speeches, and press conference transcripts offer rich linguistic cues that can influence market expectations and policy interpretation~\cite{blinder2008central, fortes2020tracking}. The evolution of computational linguistics has made it possible to extract patterns, narratives, and sentiment embedded within these documents~\cite{cecchetti2020monetary}.
	
	\subsection{Sentiment Analysis Techniques in Central Bank Communication}
	Lexicon-based sentiment analysis, particularly using the Loughran--McDonald dictionary~\cite{loughran2011liability}, was among the first attempts to quantify tone in financial texts. This approach categorizes words as positive, negative, uncertain, or litigious based on financial context. Later empirical studies~\cite{jegadeesh2015word, hansen2018transparency} showed that shifts in central bank tone could predict changes in interest rate expectations and asset prices. More advanced methods integrate domain-specific lexicons with traditional NLP pipelines, enhancing robustness.
	
	\subsection{Deep Learning and the Emergence of Contextual Language Models}
	With the advent of deep learning, transformer-based models like BERT and FinBERT have substantially improved contextual understanding in text data~\cite{araci2019finbert}. FinBERT, fine-tuned on financial corpora, excels in extracting nuanced sentiment and semantic features from central bank documents. These models outperform earlier techniques in capturing subtleties such as policy uncertainty, indirect signaling, and sentiment asymmetry. However, most studies employing these models treat text in isolation, without integrating structured macroeconomic signals.
	
	\subsection{Toward Multimodal Prediction}
	The integration of structured and unstructured data sources is still nascent. Some recent studies attempt to fuse time series data with text embeddings, using concatenation or attention mechanisms~\cite{wong2025portfolio}. Yet, comprehensive frameworks for monetary policy prediction remain limited. Bridging this gap requires not only technical innovation but also interpretability to ensure insights are meaningful to economists and policymakers. Our approach contributes to this space by unifying sentiment features with macroeconomic indicators, enabling a more holistic and explainable forecast.

	\section{Description of Dataset}
	
	This section outlines the data sources used in our multi-modal forecasting framework. The dataset comprises both structured macroeconomic indicators and unstructured textual data from official Federal Reserve communications. The target variable is constructed from historical interest rate policy decisions made by the Federal Open Market Committee (FOMC).
	
	\subsection{Structured Dataset: Economic Indicators}
	
	The structured component consists of key macroeconomic indicators commonly referenced in monetary policy analysis. These include:
	
	\begin{itemize}
		\item Inflation measures: Consumer Price Index (CPI), Personal Consumption Expenditures (PCE) price index
		\item Labor market indicators: Unemployment rate, Nonfarm Payroll Employment (NFP)
		\item Housing metrics: Housing Starts (HOUST), Home Price Index (HPI)
		\item Interest rate spreads: 10-Year Treasury Yield minus 3-Month Treasury Bill (10Y--3M spread)
	\end{itemize}
	
	All data are retrieved from the Federal Reserve Economic Data (FRED) database, maintained by the Federal Reserve Bank of St. Louis~\cite{fred2025}. Features are transformed into monthly and year-over-year differences where appropriate and standardized to ensure comparability across input dimensions.
	
	\subsection{Unstructured Dataset: Federal Reserve Communications}
	
	The unstructured dataset includes official textual releases by the Board of Governors of the Federal Reserve System. These documents span from January 2011 to January 2025 and cover five key types of communication:
	
	\begin{itemize}
		\item FOMC Statements
		\item Meeting Minutes
		\item Speeches by Federal Reserve officials
		\item Testimonies before Congress
		\item Press Conference transcripts (presconf), which represent the prepared remarks delivered by the Federal Reserve Chair
	\end{itemize}
	
	Text preprocessing involves tokenization, stopword removal, and lemmatization. Sentiment features are extracted using two approaches: (1) Term Frequency--Inverse Document Frequency (TF--IDF) vectors combined with Loughran--McDonald sentiment scores~\cite{loughran2011liability}, and (2) sentiment class probabilities generated by FinBERT, a transformer-based language model fine-tuned for financial text classification~\cite{araci2019finbert}.
	
	\subsection{Target Variable}
	
	The prediction task is framed as a three-class classification problem, with the target variable representing the direction of FOMC interest rate decisions. Each policy action is labeled as one of the following:
	
	\begin{itemize}
		\item \textit{Raise} -- indicating an increase in the federal funds target rate
		\item \textit{Hold} -- indicating no change in the target rate
		\item \textit{Lower} -- indicating a decrease in the target rate
	\end{itemize}
	
	Labels are derived from official post-meeting rate announcements and validated against historical FOMC decision records published by the Federal Reserve~\cite{board2024fomc}. This formulation enables a structured evaluation of how well each data modality contributes to the forecast of monetary policy shifts.
	
	\section{Baseline Models}
	
	This section presents the construction, training, and evaluation of baseline models that establish performance benchmarks for monetary policy classification. The methodology spans four key components: data preprocessing, feature engineering, model selection, and evaluation design.
	
	\subsection{Data Preprocessing and Feature Engineering}
	
	We begin by assembling a unified dataset that integrates structured macroeconomic indicators with unstructured textual sentiment from Federal Reserve communications. Macroeconomic variables—such as inflation, employment, and GDP growth—are standardized to ensure temporal coherence and comparability across features. 
	
	To capture linguistic signals, we extract sentiment scores from monetary policy documents using FinBERT, a domain-specific transformer model fine-tuned for financial text. Sentiment probabilities (positive, negative, and neutral) are aggregated at both the document and policy decision levels. In parallel, TF–IDF representations are constructed for each communication, and Loughran–McDonald sentiment scores are appended to enrich interpretability. Collectively, these features form a comprehensive pipeline capturing both economic fundamentals and narrative tone.
	
	\subsection{Model Architecture}
	
	We benchmark several supervised learning algorithms to establish comparative performance baselines. These include Logistic Regression, Random Forest, Extra Trees, and Gradient Boosting classifiers. For text-based inputs, we consider two representations: (i) traditional TF–IDF embeddings enhanced with sentiment scores, and (ii) FinBERT-derived sentiment probabilities. For hybrid models, we concatenate these representations with structured macroeconomic variables to assess joint predictive strength. Among these, Gradient Boosting consistently demonstrated superior capability in modeling complex and nonlinear feature interactions.
	
	\subsection{Experimental Design}
	
	To ensure rigorous and generalizable evaluation, we adopt a stratified 5-fold cross-validation strategy. This preserves label distribution across splits and mitigates overfitting. Performance is assessed using both threshold-independent metrics (ROC AUC) and threshold-dependent metrics (accuracy, precision, recall, and F1-score). Given the class imbalance in the target variable—particularly underrepresentation of “Hike” and “Cut” decisions—we incorporate the Synthetic Minority Oversampling Technique (SMOTE) and class-weighted loss functions during training. This strategy reflects real-world challenges in imbalanced policy classification.
	
	\subsection{Model Tuning and Class Imbalance Handling}
	
	To optimize model performance, we employ a two-stage tuning strategy. First, we perform randomized hyperparameter search to efficiently navigate the parameter space, followed by fine-tuning using grid search. For the Gradient Boosting classifier, the optimal configuration includes:
	
	\begin{itemize}
		\item \texttt{n\_estimators = 10}
		\item \texttt{learning\_rate = 0.01}
		\item \texttt{max\_depth = 4}
		\item \texttt{max\_features = 'sqrt'}
		\item \texttt{min\_samples\_leaf = 10}
		\item \texttt{min\_samples\_split = 10}
	\end{itemize}
	
	Four classifiers are benchmarked under this framework. Each model is trained using stratified cross-validation, and its generalization capacity is assessed on a held-out test set. The class imbalance problem is further addressed by integrating SMOTE, which synthetically augments minority class instances to promote equitable learning.

	\begin{figure}[t]
		\centering
		\includegraphics[width=0.95\linewidth]{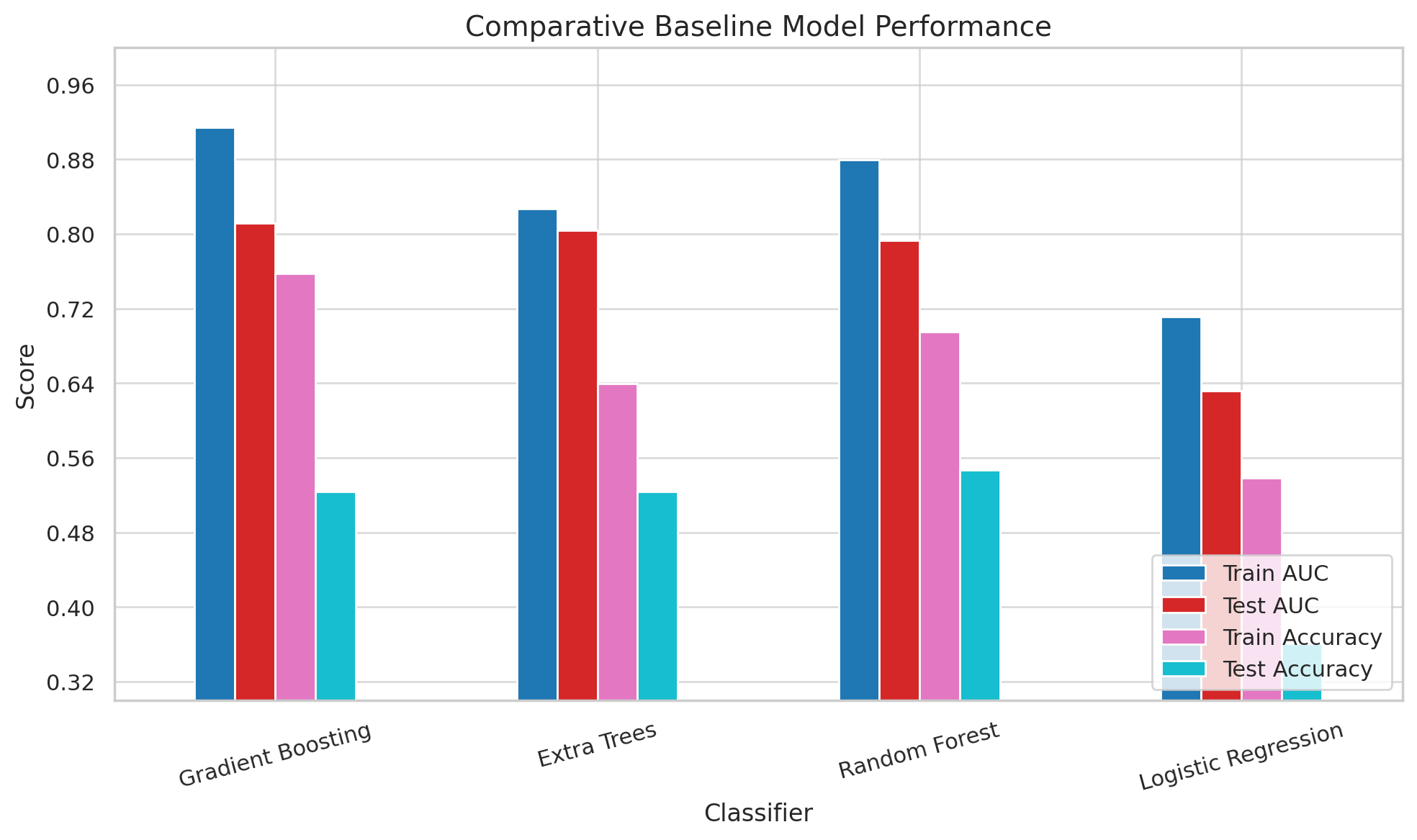}
		\caption{Baseline comparison of classifier performance in terms of ROC AUC and accuracy. Gradient Boosting consistently outperforms other models on both training and test sets. Logistic Regression exhibits underfitting, while tree-based models demonstrate stronger predictive capacity.}
		\label{fig:baseline_comparison}
	\end{figure}
	
	\subsection{Performance Summary and Insights}
	
	As illustrated in Figure~\ref{fig:baseline_comparison}, gradient Boosting delivers the most robust results across both AUC and accuracy metrics. It effectively captures complex feature interactions and remains resilient under cross-validation. Extra Trees and Random Forest models perform moderately well but show tendencies toward overfitting. Logistic Regression, while interpretable, lags in generalization and fails to capture the nuances of multimodal inputs.
	
	Our tuned Gradient Boosting model achieves excellent in-sample performance (Train AUC: 0.9139; Accuracy: 75.7\%) and respectable generalization on the test set (Test AUC: 0.8116; Accuracy: 52.3\%). Notably, after applying SMOTE to mitigate class imbalance, the model’s AUC improved, though accuracy declined slightly due to the introduction of synthetic variance.
	
	In conclusion, Gradient Boosting emerges as a strong candidate for further development within our multimodal prediction framework. Its balance between predictive power and robustness makes it well-suited for capturing the intricacies of monetary policy classification, driven by both data and discourse.

	\section{Text-Only Models}
	
	This section investigates the standalone predictive power of unstructured textual data drawn from official Federal Reserve communications. Specifically, it evaluates whether policy-related texts—such as FOMC statements, meeting minutes, press conferences, and speeches—contain sufficient informational signals to forecast U.S. interest rate decisions without the aid of traditional economic indicators.
	
	We adopt a rigorous pipeline consisting of natural language preprocessing, exploratory linguistic and sentiment analyses, and the application of both traditional machine learning classifiers and transformer-based deep learning models. The objective is to assess the quality and limitations of textual signals in capturing monetary policy intent.
	
	\subsection{Data Preprocessing}
	
	To prepare the unstructured text data for modeling, a standardised cleaning process was applied across all document types. This included case normalization, removal of extraneous characters, punctuation filtering, and stopword removal. Importantly, lemmatization was excluded to preserve domain-specific terminology critical in financial discourse.
	
	Each document was aligned with the Federal Reserve's corresponding rate decision using the FOMC calendar. In cases of overlapping or compound documents, chunking strategies were employed to handle input length constraints for transformer models.

	\subsection{Exploratory Data Analysis}
	
	We conducted exploratory linguistic analyses to uncover patterns across different document types and decision categories. Distributions of word counts were visualized to inform model constraints (e.g., BERT's token limits), and top-words analysis revealed vocabulary clusters associated with each decision type—offering early evidence of tone divergence between hawkish, neutral, and dovish communications.
	
\begin{figure}[t]
		\centering
		\includegraphics[width=\linewidth]{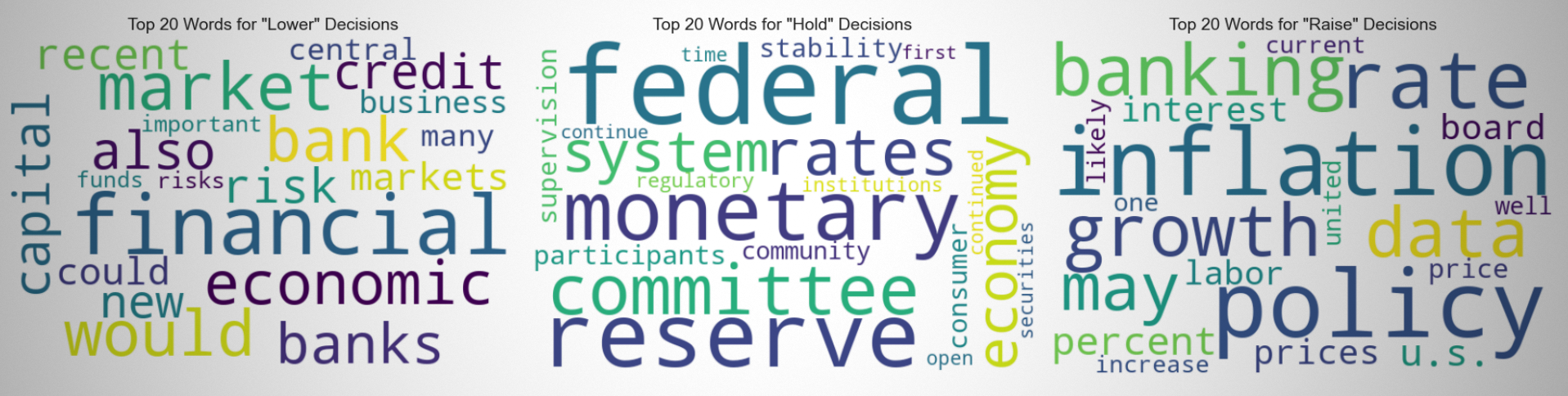}
		\caption{Top words per decision class. Themes vary significantly across monetary policy stances.}
		\label{fig2_wordclouds}
	\end{figure}

	We next analyze dominant terms associated with each decision class: \textit{Lower}, \textit{Hold}, and \textit{Raise}. As shown in Figure~\ref{fig2_wordclouds}, references to financial instability dominate the language of "Lower" decisions, while "Raise" statements emphasize inflation and labor markets. "Hold" decisions reflect a neutral, status-quo tone.
	
	\subsection{Sentiment Analysis}
	
	To quantify linguistic tone, we applied both rule-based and model-based sentiment methods.
	
	\subsubsection{Dictionary-Based Sentiment Analysis}
	
\begin{figure}[t]
		\centering
		\includegraphics[width=\linewidth]{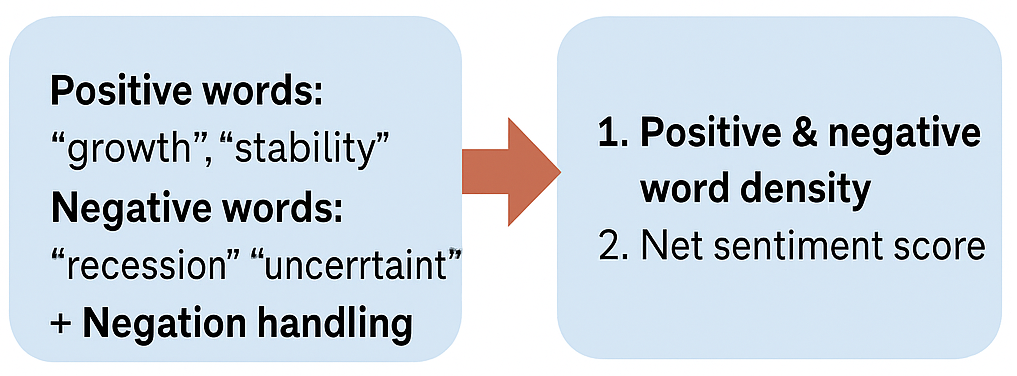}
		\caption{Rule-based sentiment pipeline using Loughran--McDonald dictionary.}
		\label{fig3_lm_pipeline}
	\end{figure}

	Using the Loughran–McDonald (LM) sentiment lexicon \cite{loughran2011liability}, we computed sentiment scores from positive and negative word counts, normalized by document length. A negation-aware scoring mechanism was implemented to reduce semantic errors. We extracted three metrics: positive density, negative density, and net sentiment. Temporal and categorical analyses showed that sentiment shifts generally align with known macroeconomic stress periods, validating the LM dictionary's utility in formal policy text.	Figure~\ref{fig3_lm_pipeline} summarizes this extraction pipeline.
	
\begin{figure}[t]
		\centering
		\includegraphics[width=\linewidth]{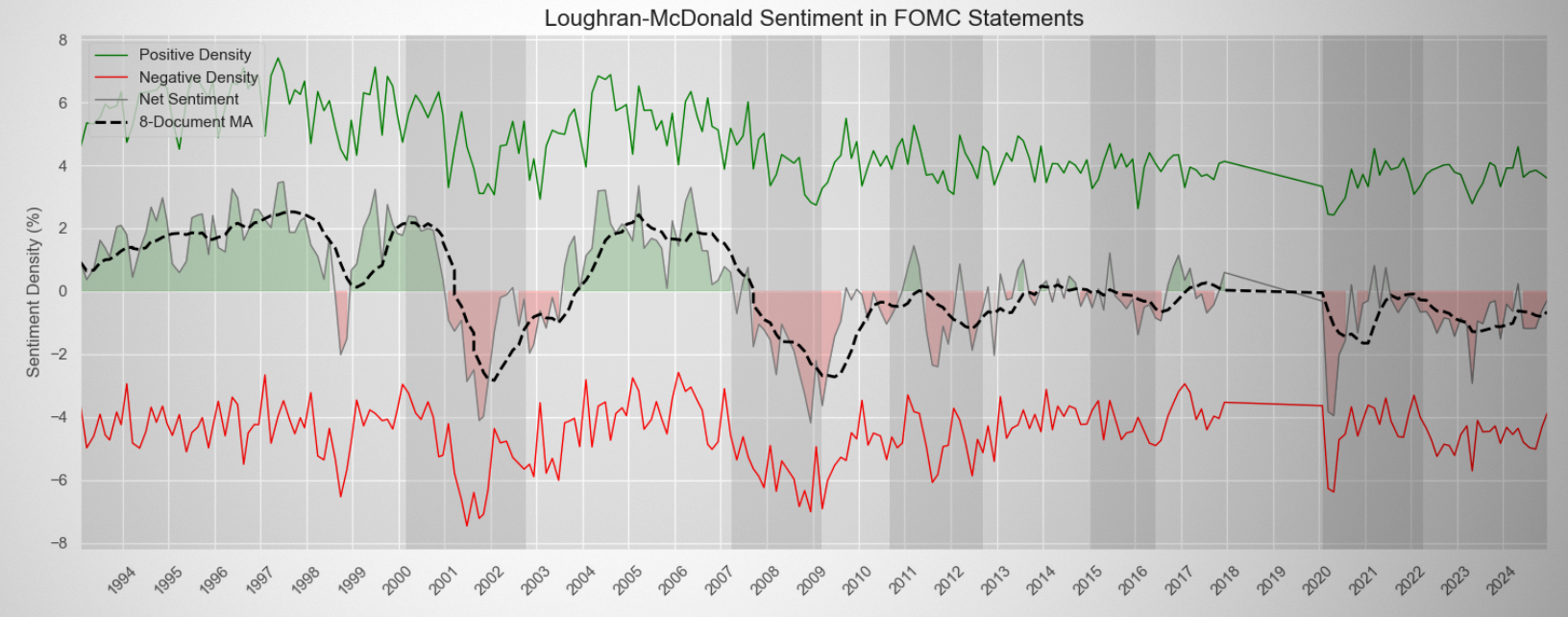}
		\caption{Net sentiment trends over time. Dips align with U.S. recession periods (shaded).}
		\label{fig4_sentiment_cycles}
	\end{figure}

	Figure~\ref{fig4_sentiment_cycles} presents the evolution of net sentiment over time. Notably, sentiment declines tend to precede major recession periods, reinforcing its potential as a leading indicator.
	
	\begin{figure}[t]
		\centering
		\includegraphics[width=0.5\textwidth]{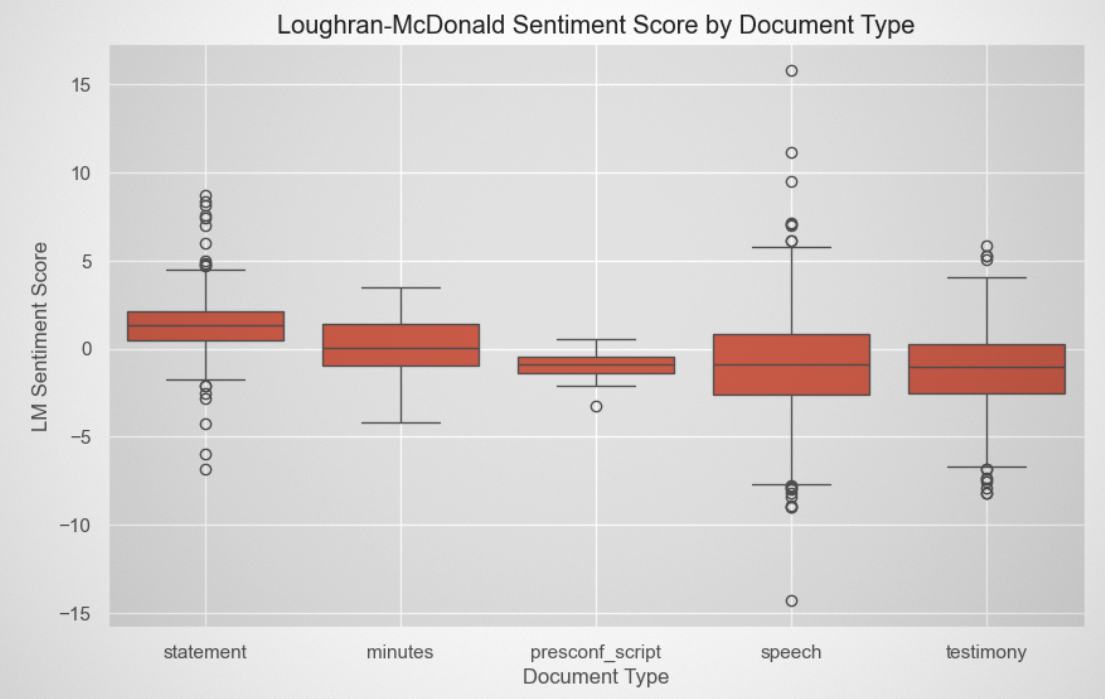}
		\caption{Distributions of LM sentiment scores by document type.}
		\label{fig:lm_sentiment_distribution}
	\end{figure}
	
	As seen in Figure~\ref{fig:lm_sentiment_distribution}, sentiment polarity varies by document type. Statements show a slightly positive skew, whereas speeches and testimonies present more diverse and often negative sentiment. Press conferences (\textit{presconf}) exhibit a narrow, neutral range, aligning with their explanatory tone. These distinctions offer valuable inputs for forecasting interest rate
	decisions.
	
	\subsubsection{Transformer-Based Sentiment Analysis}
	
	We also employed FinBERT\cite{araci2019finbert}, a domain-specific transformer model trained on financial corpora, to classify sentiment across the document corpus. Predictions were aggregated from chunk-level softmax probabilities to derive document- and decision-level sentiment summaries. While FinBERT returned mostly neutral classifications, its probabilistic outputs were retained as features in downstream models. FinBERT’s strength lies in contextual understanding, offering a probabilistic counterpoint to the interpretable yet rigid dictionary-based scores.

	\begin{figure}[t]
		\centering
		\includegraphics[width=\linewidth]{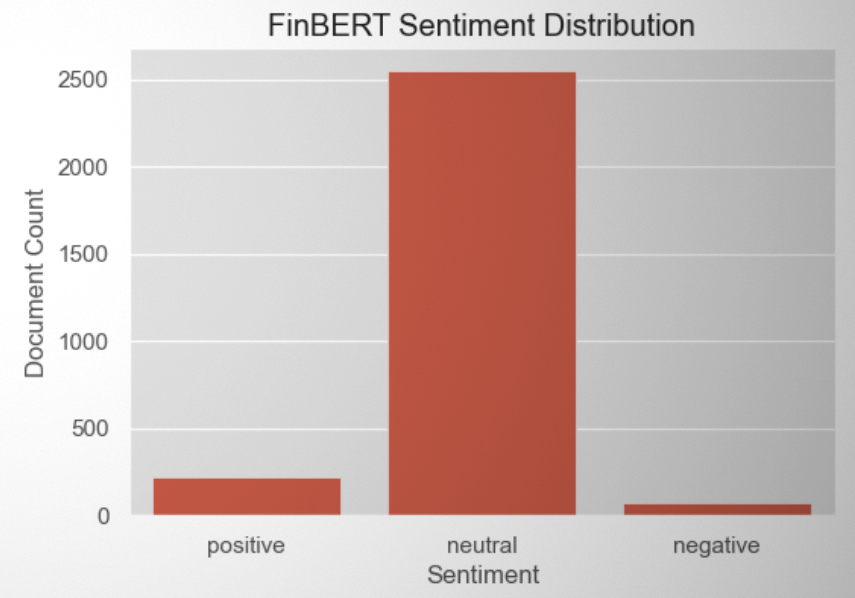}
		\caption{FinBERT sentiment classification across central bank documents. Majority are labeled as neutral.}
		\label{fig6_finbert_dist}
	\end{figure}
	
	As shown in Figure~\ref{fig6_finbert_dist}, most communications are labeled "neutral," consistent with the Fed's measured communication style. FinBERT’s sentiment outputs mirror the measured and neutral tone commonly found in central bank communications, leading to limited variation across sentiment categories. Although less effective in isolation, these scores serve as a valuable probabilistic complement to rule-based methods, motivating the exploration of FinBERT as a direct classifier to capture nuanced financial language better.

	\subsection{Model Training and Evaluation}
	
	Two categories of models were developed using only textual inputs to forecast interest rate decisions.
	
	\begin{itemize}
		\item \textbf{Traditional classifiers} (Logistic Regression, Naïve Bayes, Random Forest, Extra Trees, Gradient Boosting) were trained on TF-IDF representations of cleaned texts. Despite some variation in accuracy, overall AUC scores remained modest, suggesting that shallow models struggle to extract deep context from policy language.
		\item \textbf{FinBERT fine-tuned classifier} was trained to directly predict the rate decision class. Document chunking and weighted cross-entropy loss were applied to address token constraints and class imbalance. The best checkpoint achieved a validation ROC AUC of 0.69 and accuracy of 0.67, outperforming traditional models but still showing a persistent bias toward the majority “Hold” class.
	\end{itemize}
	
	\subsection{Evaluation Insights}
	
	Although textual models offer some predictive capacity—especially with FinBERT’s contextual strengths—the performance ceiling appears constrained by the formal, neutral tone of Fed communications. Discriminative signals are subtle and often insufficient when used in isolation. As summarized in Table~\ref{tab:text_only_results}, the standalone use of unstructured text falls short of capturing the full complexity of interest rate decisions. These limitations motivate the integration of structured economic data into a multi-modal learning framework.
	
	\begin{table}[t]
		\centering
		\caption{Text-only models predict interest rate decisions}
		\label{tab:text_only_results}
		\begin{tabular}{lcc}
			\toprule
			\textbf{Model} & \textbf{Test ROC AUC} & \textbf{Test Accuracy} \\
			\midrule
			TF-IDF + Logistic Regression        & 0.6290 & 0.4959 \\
			TF-IDF + Gradient Boosting          & 0.6759 & 0.5190 \\
			FinBERT (fine-tuned)       & \textbf{0.6869} & \textbf{0.6690} \\
			\bottomrule
		\end{tabular}
	\end{table}
	
	\section{Multi-Modal Models}
	
	To explore how structured economic data (ED) and unstructured data can complement each other, we develop three multi-modal modeling frameworks. Each framework combines macroeconomic indicators with a different form of textual representation. Our goal is to understand how different combinations influence the accuracy and interpretability of monetary policy forecasts.
	
	\subsubsection{Method 1: ED + TF-IDF and LM Sentiment Features + XGBoost}

	\begin{algorithm}
		\caption{Combine TF-IDF and LM Sentiment Features}
		\label{alg:tfidf_lm_combination}
		\begin{algorithmic}[1]
			\Require Document corpus $D = \{d_1, d_2, \dots, d_n\}$
			\Require LM sentiment lexicon $L = \{L_{\text{pos}}, L_{\text{neg}}, L_{\text{unc}}, L_{\text{lit}}\}$
			
			\State Preprocess each document: remove stopwords, lowercase, tokenize
			\For{each document $d_i$ in $D$}
			\State Compute TF-IDF vector $T_i \gets \text{TFIDF}(d_i)$
			\State Initialize sentiment counts: $s_{\text{pos}}, s_{\text{neg}}, s_{\text{unc}}, s_{\text{lit}} \gets 0$
			\For{each token $t$ in $d_i$}
			\If{$t \in L_{\text{pos}}$} \State $s_{\text{pos}} \gets s_{\text{pos}} + 1$ \EndIf
			\If{$t \in L_{\text{neg}}$} \State $s_{\text{neg}} \gets s_{\text{neg}} + 1$ \EndIf
			\If{$t \in L_{\text{unc}}$} \State $s_{\text{unc}} \gets s_{\text{unc}} + 1$ \EndIf
			\If{$t \in L_{\text{lit}}$} \State $s_{\text{lit}} \gets s_{\text{lit}} + 1$ \EndIf
			\EndFor
			\State Normalize sentiment scores by total tokens if desired
			\State Form LM sentiment vector $L_i \gets [s_{\text{pos}}, s_{\text{neg}}, s_{\text{unc}}, s_{\text{lit}}]$
			\State Concatenate: $F_i \gets [T_i, L_i]$
			\EndFor
			
			\State \Return Feature matrix $F = \{F_1, F_2, \dots, F_n\}$ for model training
		\end{algorithmic}
	\end{algorithm}

	The first framework merges structured economic features with two sets of textual inputs:  
	(1) 500-dimensional TF-IDF vectors derived from Federal Reserve communications, and  
	(2) 50 sentiment features based on the Loughran–McDonald (LM) financial dictionary.
	
	\noindent
	Algorithm~\ref{alg:tfidf_lm_combination} outlines the process for combining TF-IDF and Loughran–McDonald (LM) sentiment features into a unified feature matrix for model training. The input consists of a document corpus and the LM sentiment lexicon, which includes predefined lists of positive, negative, uncertain, and litigious words. Each document is first preprocessed through tokenization, stopword removal, and lowercasing. Next, a TF-IDF vector is computed to capture the importance of terms in the corpus. Simultaneously, sentiment counts are tallied by matching tokens against the LM lexicon categories. These counts are optionally normalized and then concatenated with the TF-IDF vector to form a combined feature representation. The resulting feature matrix integrates both frequency-based textual relevance and domain-specific sentiment signals, enabling richer input for downstream classification or prediction tasks.

	\begin{figure}[t]
		\centering
		\includegraphics[width=0.45\textwidth]{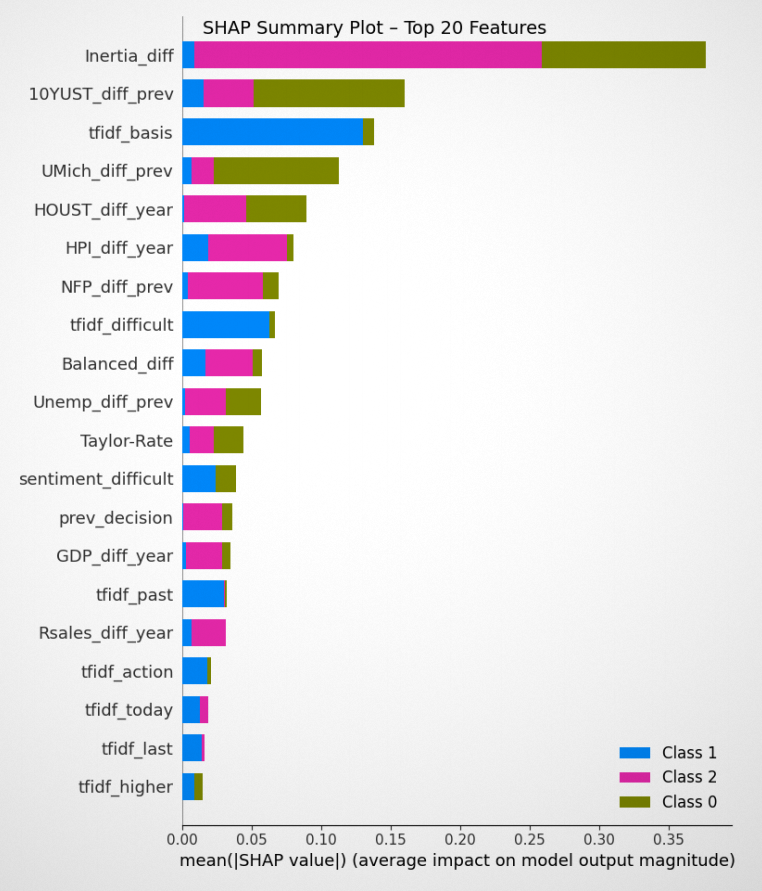}
		\caption{Method 1: SHAP summary plot highlighting the top features influencing the XGBoost model’s predictions}
		\label{fig:shap_xgboost_method1}
	\end{figure}

	\begin{figure}[h]
		\centering
		\includegraphics[width=0.45\textwidth]{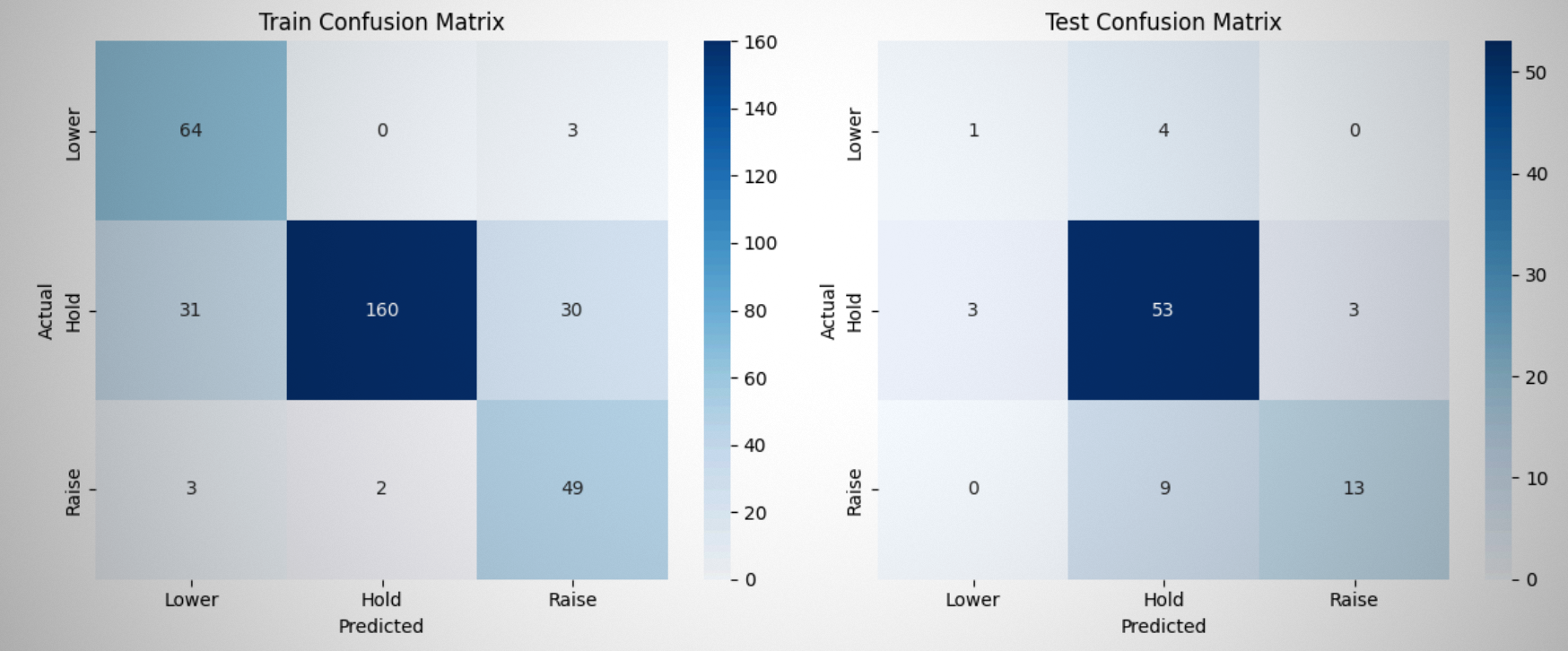}
		\caption{Method 1: ED + TF-IDF \& LM Sentiment + XGBoost}
		\label{fig:xgboost_cm_tfidf}
	\end{figure}
	
	We use these combined features to train an XGBoost classifier. Figure~\ref{fig:shap_xgboost_method1} shows the SHAP summary plot for the XGBoost model using economic and textual features. The most influential variable is the deviation from policy inertia (\texttt{Inertia\_diff}), which captures intentional shifts in monetary policy behavior. Bond market expectations (\texttt{10YUST\_diff\_prev}) and consumer sentiment (\texttt{UMich\_diff\_prev}) also exhibit strong contributions, especially in identifying “Lower” rate decisions. Housing-related indicators (\texttt{HOUST\_diff\_year}, \texttt{HPI\_diff\_year}) reflect broader macroeconomic conditions and play a notable role. Additionally, textual features such as \texttt{tfidf\_basis} and \texttt{tfidf\_difficult}, extracted from FOMC statements, capture subtle narrative tones and help improve both prediction accuracy and model interpretability. This distribution of feature importance supports the value of combining structured data with domain-specific textual cues.

	Figure~\ref{fig:xgboost_cm_tfidf} shows the confusion matrix for Method 1. The model performs well across all three decision categories—“Raise,” “Hold,” and “Lower.” Most predictions fall along the diagonal, indicating a high level of agreement with actual FOMC outcomes. While some confusion exists between “Hold” and the adjacent classes, the overall balance suggests that the combined feature set captures relevant policy signals effectively. The model is particularly accurate in identifying the most frequent “Hold” decisions, without severely misclassifying the less common classes.
	
	This model performs strongly, achieving a test AUC of \textbf{0.8304} and an accuracy of \textbf{77.91\%}. Its balance of simplicity and interpretability makes it a promising baseline.
	
	\subsubsection{Method 2: ED + FinBERT Sentiment Probabilities + XGBoost}
	
	In the second modeling framework, we integrated structured economic indicators with sentiment classification probabilities generated by a fine-tuned FinBERT model. While the \texttt{TF-IDF} approach captures surface-level word frequencies, this transformer-based method was designed to extract deeper contextual meaning from Federal Reserve communications. The goal was to evaluate whether FinBERT's domain-specific sentiment scores could enhance predictive performance when combined with macroeconomic data.
	
\begin{figure}[t]
		\centering
		\includegraphics[width=0.45\textwidth]{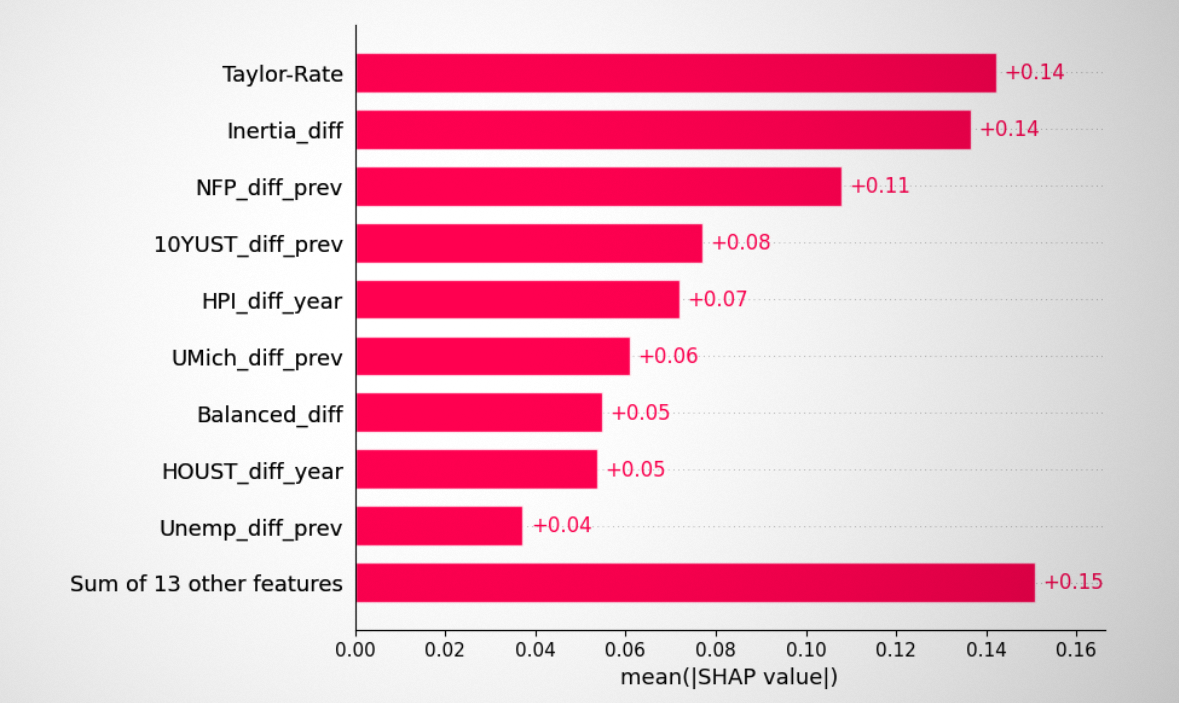}
		\caption{Method 2: SHAP Summary Plot}
		\label{fig:shap_xgboost_method2}
	\end{figure}

	We trained an \texttt{XGBoost} classifier on the combined feature set. As illustrated in the SHAP summary plot (Figure~\ref{fig:shap_xgboost_method2}), macroeconomic variables remained the most influential predictors. These included the Taylor Rule-implied interest rate (\texttt{Taylor\_Rate}), the deviation from policy inertia (\texttt{Inertia\_diff}), labor market momentum (\texttt{NFP\_diff\_prev}), bond yield changes (\texttt{10YUST\_diff\_prev}), and consumer sentiment (\texttt{UMich\_diff\_prev}).
	
	By contrast, the FinBERT-derived sentiment probabilities contributed modestly to the model's decision process. This suggests that although \texttt{FinBERT} captures rich contextual nuance, the relatively shallow architecture of \texttt{XGBoost} may not fully exploit its representational depth. These results highlight a trade-off between model simplicity and the complexity of text-based features in financial policy prediction tasks.

	\begin{figure}[t]
		\centering
		\includegraphics[width=0.45\textwidth]{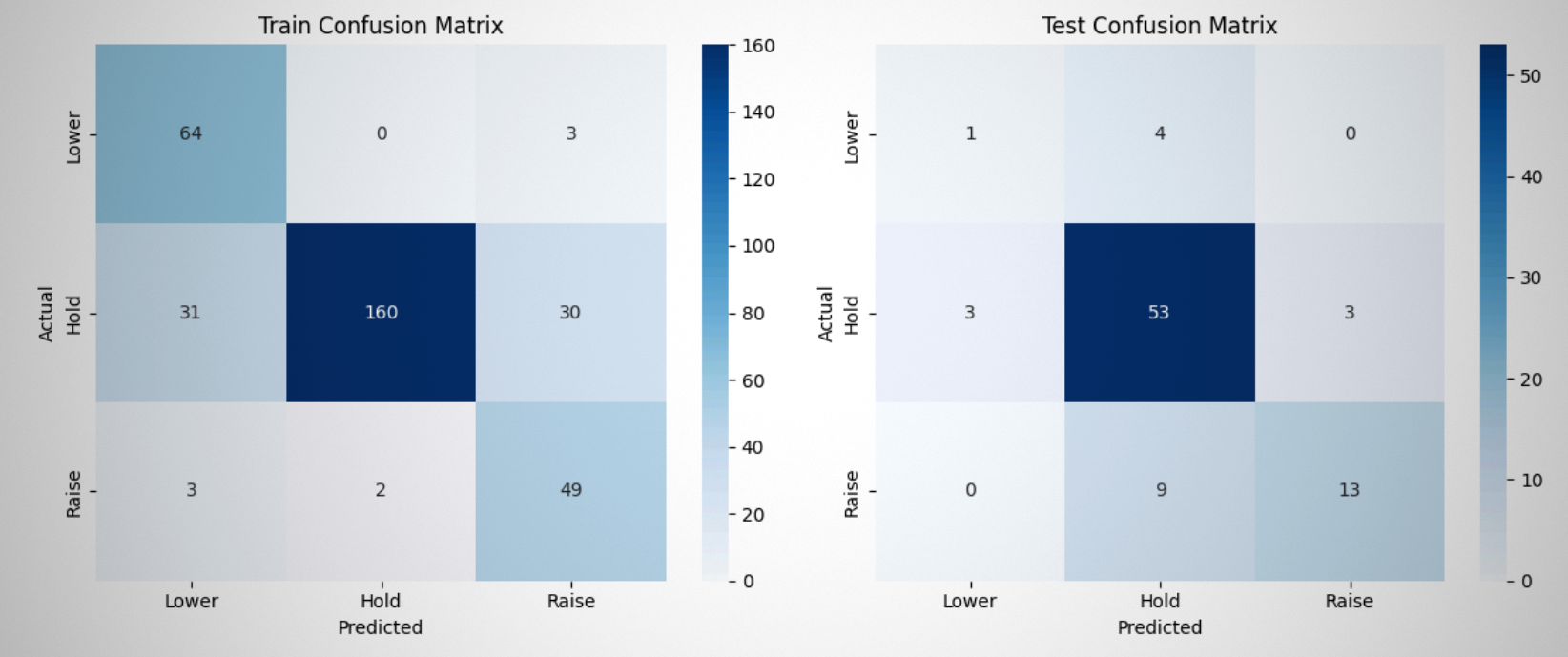}
		\caption{Method 2: ED + FinBERT Sentiment + XGBoost}
		\label{fig:xgboost_cm_finbert}
	\end{figure}

	Figure~\ref{fig:xgboost_cm_finbert} presents the confusion matrix for Method 2. The model shows reasonable performance in predicting “Hold” decisions but struggles with minority classes, especially “Raise.” Many instances of “Raise” are incorrectly predicted as “Hold,” highlighting the challenge posed by class imbalance. This pattern also suggests that FinBERT sentiment probabilities, while capturing general tone, may lack the precision needed to differentiate more subtle policy shifts. Compared to Method 1, this approach trades off some classification accuracy for sentiment-based contextual richness.
	
	The model achieves a test AUC of \textbf{0.7960} and an accuracy of \textbf{59.30\%}. Performance declines suggest that sentiment probabilities from FinBERT may smooth over subtle textual distinctions. The effect is more noticeable under class imbalance.
	
	\subsubsection{Method 3: ED + FinBERT Sentiment Probabilities + FNN}
	
	The third framework tests a deep learning setup. We use a feedforward neural network (FNN) with two hidden layers (64 and 32 units). Input features include economic indicators and FinBERT sentiment probabilities.
	
	\begin{figure}[t]
		\centering
		\includegraphics[width=0.45\textwidth]{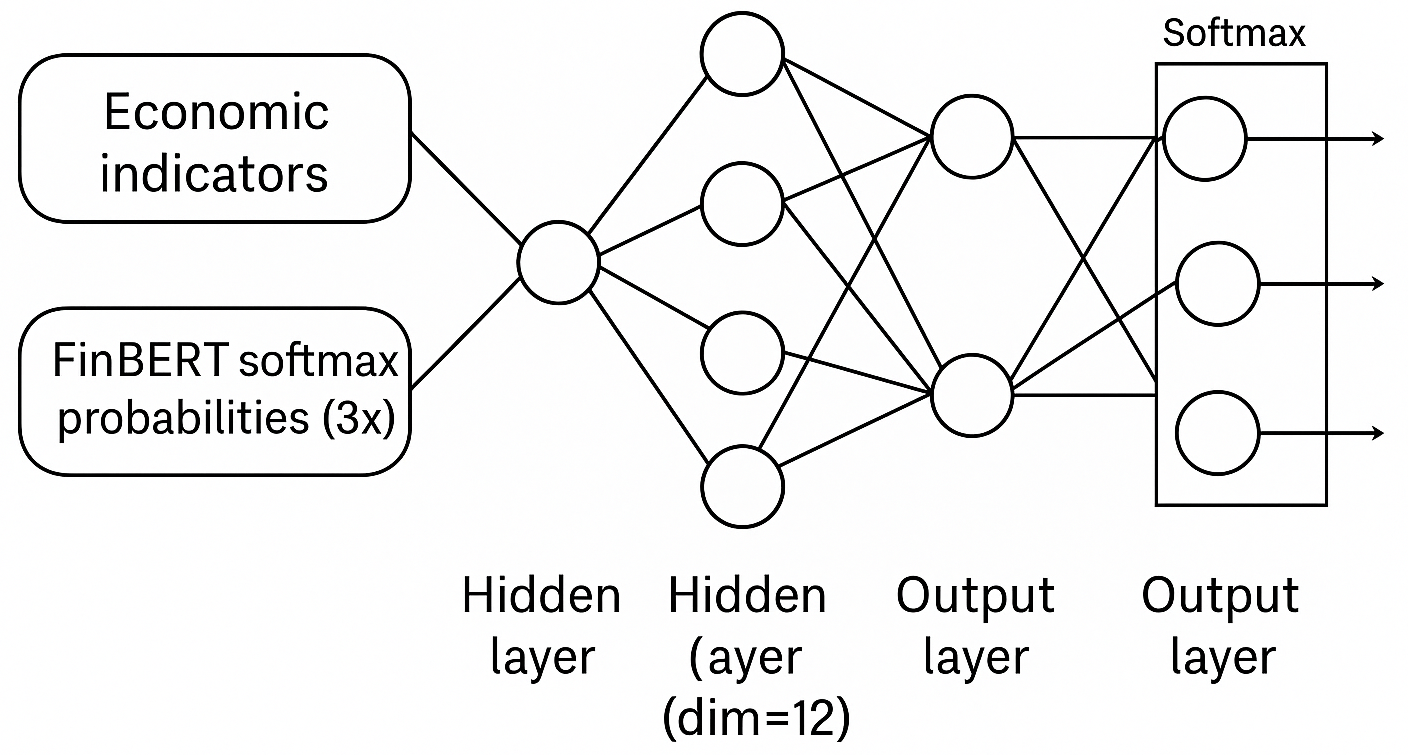}
		\caption{Feedforward Neural Network (FNN) Architecture}
		\label{fig:fnn_architecture}
	\end{figure}
	
	\begin{figure}[t]
		\centering
		\includegraphics[width=0.45\textwidth]{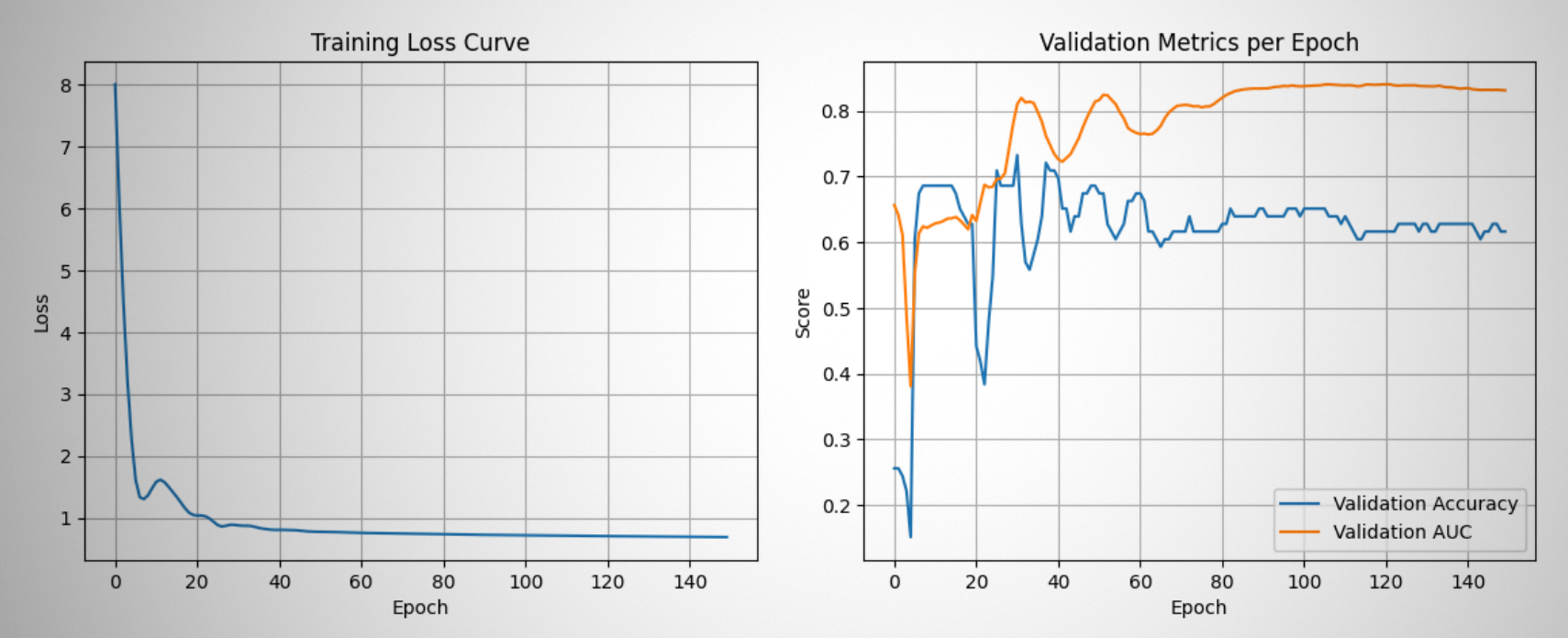}
		\caption{FNN Training Curve: Loss, AUC, and Accuracy}
		\label{fig:fnn_metrics}
	\end{figure}
	
	\begin{figure}[t]
		\centering
		\includegraphics[width=0.3\textwidth]{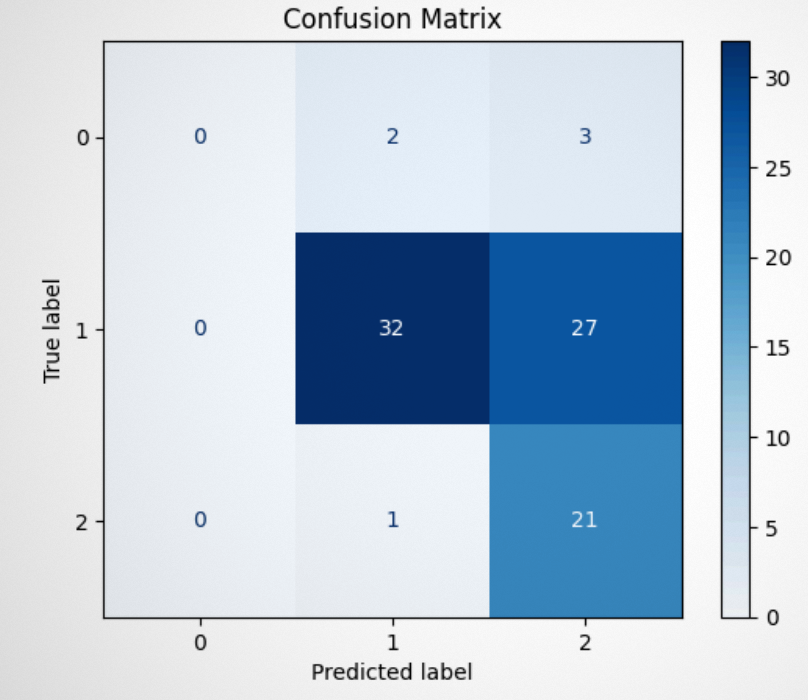}
		\caption{Method 3: ED + FinBERT Sentiment + FNN}
		\label{fig:stage3_confusion_metrics}
	\end{figure}
	
	This framework achieves the highest test AUC at \textbf{0.8404}, suggesting strong ranking ability. However, its accuracy is lower at \textbf{61.63\%}. The FNN struggles with calibration and misclassifies less common classes like “Raise” and “Lower.” Sensitivity to class imbalance may explain this behavior.
	
	\subsubsection{Comparative Performance and Insights}
	
	\begin{figure}[t]
		\centering
		\includegraphics[width=0.45\textwidth]{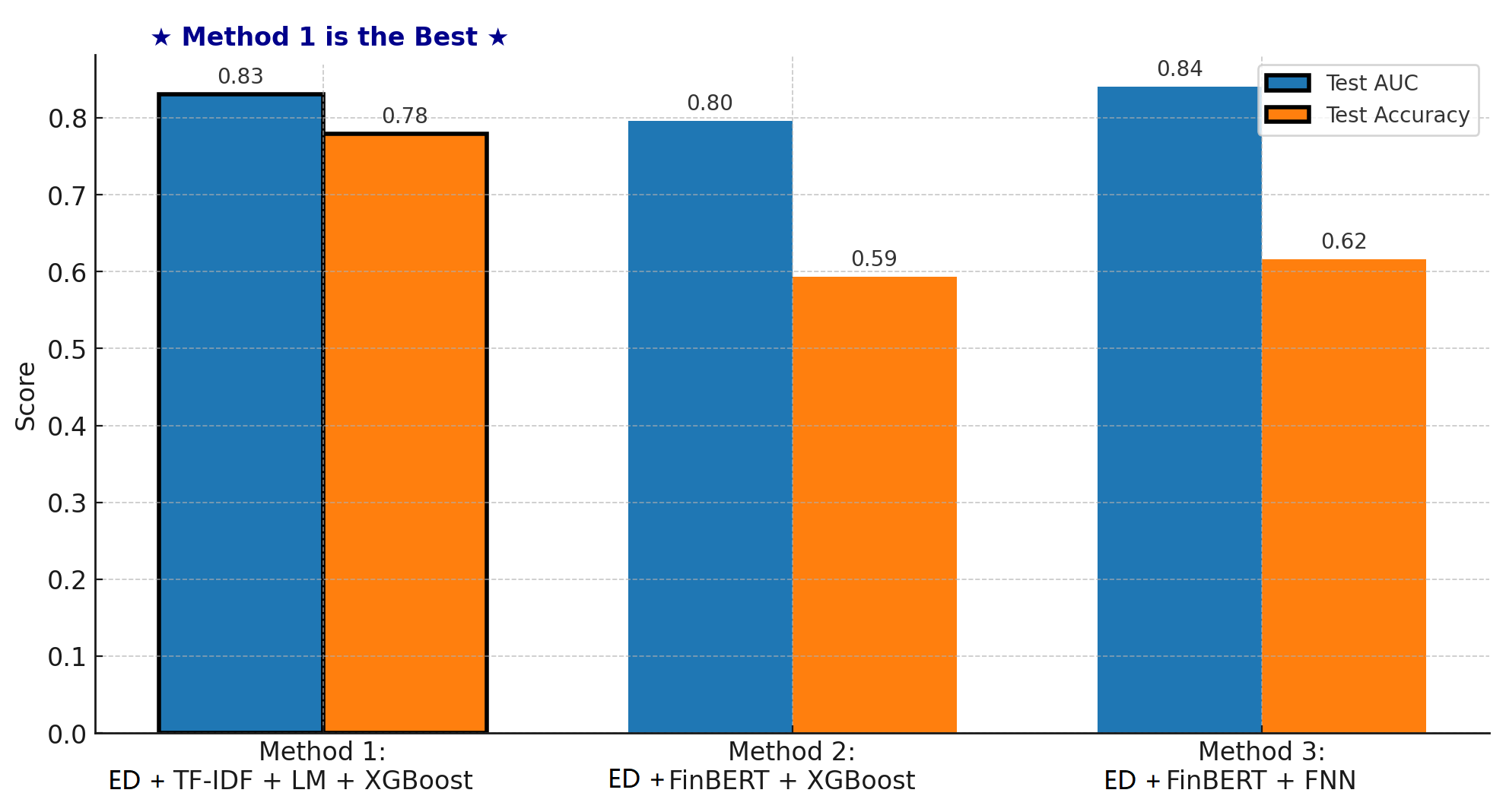}
		\caption{Comparative Performance of Hybrid Models}
		\label{fig:hybrid_performance}
	\end{figure}
	
	Figure~\ref{fig:hybrid_performance} compares all three hybrid models. The FNN shows the highest AUC but lower overall accuracy. In contrast, the XGBoost model with TF-IDF and LM sentiment offers both solid performance and strong interpretability.
	
	Several insights emerge. First, sparse, transparent text features may capture formal monetary language better than dense transformer outputs. Second, simpler models like XGBoost remain competitive—particularly when paired with carefully selected and engineered inputs. Finally, interpretability tools like SHAP help validate economic meaning and ensure the model aligns with domain intuition.
	
	In sum, we find that hybrid models hold great promise. While deep learning offers flexibility, structured approaches built on interpretable features may be more practical for financial policy forecasting, especially when clarity and trust are essential.

	\section{Conclusion}
	
	\subsection{Key Insights}
	
	This study offers a data-driven perspective on forecasting U.S. monetary policy using multi-modal machine learning models. We examined how combining structured economic indicators with unstructured central bank communication can enhance predictive performance and interpretability. Several key insights emerged:
	
	\begin{itemize}
		\item \textbf{Integrating structured and unstructured inputs} consistently improved performance over single-modality models. This highlights the complementary value of quantitative fundamentals and qualitative narratives in policy forecasting.
		
		\item \textbf{TF-IDF features}, when combined with Loughran–McDonald sentiment scores, captured meaningful linguistic cues from FOMC documents. These sparse, interpretable signals outperformed transformer-based sentiment embeddings in both accuracy and clarity.
		
		\item \textbf{Shallow models like XGBoost}, when paired with thoughtfully engineered hybrid features, achieved the best balance of performance, robustness, and interpretability. They also managed class imbalance more effectively than deep neural networks, especially in settings with limited or skewed data.
	\end{itemize}
	
	\subsection{Limitations}
	
	We remain mindful of the limitations of our current approach. One major challenge is class imbalance, particularly the small number of “Lower” rate decisions. While we applied class-weighting and SMOTE to address this, synthetic oversampling introduced variance and did not significantly improve generalization.
	
	Additionally, FinBERT sentiment probabilities, while rich in context, lacked the granularity needed for fine-grained prediction. Their compressed format limited both precision and transparency in downstream tasks.
	
	\subsection{Future Directions}
	
	There are several promising directions for future work:
	
	\begin{enumerate}
		\item Develop targeted or ensemble classifiers to improve recall on minority classes and reduce model bias.
		\item Explore lighter, fine-tuned transformer models adapted specifically to monetary policy language.
		\item Incorporate external signals—such as market-based expectations or global economic trends—to broaden the model’s situational awareness.
	\end{enumerate}
	
	\noindent In closing, we believe this work provides a modest contribution to the evolving field of data-driven policy analysis. Our findings highlight the potential of hybrid, interpretable frameworks to support deeper understanding of central bank decision-making. While challenges remain, we hope this research encourages further exploration at the intersection of machine learning, economics, and institutional communication.


	\section{acknowledgments}
	We gratefully acknowledge the National University of Singapore (NUS) for its generous support and collaborative environment, which played a pivotal role in enabling this research. In particular, we thank the NUS School of Computing for funding this work through the Graduate Project Supervision Fund (SF). Their guidance, resources, and academic ecosystem have been instrumental in shaping and advancing the direction of this study.

	
	\bibliographystyle{IEEEtran}

\begin{thebibliography}{100}
		
		\bibitem{cecchetti2020monetary}
		S.~G.~Cecchetti, M.~S.~Mohanty, and F.~Zampolli, ``Monetary Policy in the Next Recession?,'' \emph{CEPR Policy Insight}, no.~103, 2020.
		
		\bibitem{blinder2008central}
		A.~S.~Blinder, M.~Ehrmann, M.~Fratzscher, J.~de Haan, and D.-J.~Jansen, ``Central Bank Communication and Monetary Policy: A Survey of Theory and Evidence,'' \emph{Journal of Economic Literature}, vol.~46, no.~4, pp.~910--945, 2008.
		
		\bibitem{fortes2020tracking}
		R.~Fortes and T.~Le~Guenedal, ``Tracking ECB's Communication: Perspectives and Implications for Financial Markets,'' \emph{Banque de France Bulletin}, no.~231, 2020.
		
		\bibitem{wong2025portfolio}
		J.~Wong and L.~Liu, ``Portfolio Optimization through a Multi-Modal Deep Reinforcement Learning Framework,'' \emph{Authorea Preprints}, 2025.
		
		\bibitem{taylor1993discretion}
		J.~B.~Taylor, ``Discretion versus Policy Rules in Practice,'' \emph{Carnegie-Rochester Conference Series on Public Policy}, vol.~39, pp.~195--214, 1993.
		
		\bibitem{bollerslev1986garch}
		T.~Bollerslev, ``Generalized Autoregressive Conditional Heteroskedasticity,'' \emph{Journal of Econometrics}, vol.~31, no.~3, pp.~307--327, 1986.
		
		\bibitem{loughran2011liability}
		T.~Loughran and B.~McDonald, ``When is a Liability Not a Liability? Textual Analysis, Dictionaries, and 10-Ks,'' \emph{Journal of Finance}, vol.~66, no.~1, pp.~35--65, 2011.
		
		
		\bibitem{fred2025}
		Federal Reserve Bank of St. Louis, ``Federal Reserve Economic Data (FRED),'' \emph{Online Resource}, 2025. [Online]. Available: \url{https://fred.stlouisfed.org/}
		
		\bibitem{araci2019finbert}
		D.~Araci, ``FinBERT: Financial Sentiment Analysis with Pre-Trained Language Models,'' \emph{arXiv preprint arXiv:1908.10063}, 2019.
		
		
		\bibitem{taylor1993discretion}
		J.~B. Taylor, ``Discretion versus policy rules in practice,'' \emph{Carnegie-Rochester Conference Series on Public Policy}, vol.~39, pp. 195--214, 1993.
		
		\bibitem{apel2012monetary}
		M.~Apel and M.~Grimaldi, ``Monetary policy decision-making, market expectations and commitment,'' \emph{Journal of Monetary Economics}, vol.~59, no.~6, pp. 601--621, 2012.
		
		\bibitem{bollerslev1986garch}
		T.~Bollerslev, ``Generalized autoregressive conditional heteroskedasticity,'' \emph{Journal of Econometrics}, vol.~31, no.~3, pp. 307--327, 1986.
		
		\bibitem{hansen2018transparency}
		S.~Hansen and M.~McMahon, ``Transparency and Deliberation in Monetary Policy,'' \emph{Econometrica}, vol.~86, no.~2, pp. 499--530, 2018.
		
		\bibitem{engle1982arch}
		R.~F. Engle, ``Autoregressive Conditional Heteroskedasticity with Estimates of the Variance of U.K. Inflation,'' \emph{Econometrica}, vol.~50, no.~4, pp. 987--1007, 1982.
		
		
		\bibitem{apel2012monetary}
		M.~Apel and M.~Grimaldi, ``Monetary policy decision-making, market expectations and commitment,'' \emph{Journal of Monetary Economics}, vol.~59, no.~6, pp. 601--621, 2012.
		
		\bibitem{bollerslev1986garch}
		T.~Bollerslev, ``Generalized autoregressive conditional heteroskedasticity,'' \emph{Journal of Econometrics}, vol.~31, no.~3, pp. 307--327, 1986.
		
		\bibitem{hansen2018transparency}
		S.~Hansen and M.~McMahon, ``Transparency and Deliberation in Monetary Policy,'' \emph{Econometrica}, vol.~86, no.~2, pp. 499--530, 2018.
		
		\bibitem{jegadeesh2015word}
		N.~Jegadeesh and D.~Wu, ``Word power: A new approach for content analysis,'' \emph{Journal of Financial Economics}, vol.~117, no.~2, pp. 371--394, 2015.
	
		\bibitem{bollerslev1986garch}
		T.~Bollerslev, ``Generalized Autoregressive Conditional Heteroskedasticity,'' \emph{Journal of Econometrics}, vol.~31, no.~3, pp. 307--327, 1986.
		
		\bibitem{loughran2011liability}
		T.~Loughran and B.~McDonald, ``When is a Liability not a Liability? Textual Analysis, Dictionaries, and 10‐Ks,'' \emph{The Journal of Finance}, vol.~66, no.~1, pp. 35--65, 2011.
		
		\bibitem{araci2019finbert}
		D.~Araci, ``FinBERT: Financial Sentiment Analysis with Pre-trained Language Models,'' \emph{arXiv preprint arXiv:1908.10063}, 2019.
		
		\bibitem{hansen2018transparency}
		S.~Hansen and M.~McMahon, ``Transparency and Deliberation in Monetary Policy,'' \emph{Econometrica}, vol.~86, no.~2, pp. 499--530, 2018.
		
	
		
	\end{thebibliography}

\end{document}